\documentclass{iopart}

\usepackage{epsfig}
\newcommand{\beq}{\begin{equation}}
\newcommand{\eeq}{\end{equation}}
\newcommand{\bes}{\begin{subequations}}
\newcommand{\ees}{\end{subequations}}
\newcommand{\bea}{\begin{eqnarray}}
\newcommand{\eea}{\end{eqnarray}}
\newcommand{\ba}{\begin{array}}
\newcommand{\ea}{\end{array}}
\newcommand{\beqn}{\begin{eqnarray*}}
\newcommand{\eeqn}{\end{eqnarray*}}

\newcommand{\f}[2]{\frac{#1}{#2}}
\newcommand{\lisa}{{\em LISA}}
\def\ii{{\rm i}}   
\def\lappreq{\!\stackrel{\scriptscriptstyle <}{\scriptscriptstyle\sim}\!}
\def\gappreq{\!\stackrel{\scriptscriptstyle >}{\scriptscriptstyle\sim}\!}

\begin{document}

\title[LISA observations of massive black hole mergers]{LISA
observations of massive black hole mergers: \\ event rates and issues
in waveform modelling}

\author{Emanuele Berti\footnote[3]{email: berti@wugrav.wustl.edu}}

\address{McDonnell Center for the Space Sciences, Department of
Physics, Washington University, St. Louis, Missouri 63130}

\begin{abstract}
The observability of gravitational waves from supermassive and
intermediate-mass black holes by the forecoming Laser Interferometer
Space Antenna (\lisa), and the physics we can learn from the
observations, will depend on two basic factors: the event rates for
massive black hole mergers occurring in the \lisa~best sensitivity
window, and our theoretical knowledge of the gravitational
waveforms. We first provide a concise review of the literature on
\lisa~event rates for massive black hole mergers, as predicted by
different formation scenarios. Then we discuss what (in our view) are
the most urgent issues to address in terms of waveform modelling. For
massive black hole binary inspiral these include spin precession,
eccentricity, the effect of high-order Post-Newtonian terms in the
amplitude and phase, and an accurate prediction of the transition from
inspiral to plunge. For black hole ringdown, numerical relativity will
ultimately be required to determine the relative quasinormal mode
excitation, and to reduce the dimensionality of the template space in
matched filtering.

\end{abstract}



\maketitle

The Laser Interferometer Space Antenna (\lisa) is being designed to
detect gravitational waves of frequency between $10^{-5}$ and
$10^{-1}$~Hz, with maximum sensitivity around $\sim
10^{-2}$~Hz. Astrophysical sources in this frequency band are usually
split into three broad classes: 1) a large background of galactic
binaries (mostly white dwarf binaries) with periods ranging from hours
to tens of seconds; 2) the ``extreme mass ratio inspirals'' (EMRIs) of
stars and black holes (BHs) of mass $M\sim 10 M_\odot$ into
supermassive black holes (SMBHs); 3) the coalescence of SMBH binaries
and the capture of intermediate-mass black holes (IMBHs) by SMBHs.

The distinction between IMBHs (with mass $M\sim 10^2-10^4~M_\odot$)
and SMBHs (with $M\sim 10^5-10^9~M_\odot$) is not sharp, and we will
refer to both of them collectively as ``massive black holes''
(MBHs). MBHs will be observed with large signal-to-noise ratio (SNR),
allowing precise measurements of the source parameters and tests of
the strong-field effects of general relativity, both in the inspiral
\cite{BBW,BBW2} and ringdown \cite{BCW,BCW2} phases. For this reason
the observation of MBH mergers is one of the top \lisa~science
milestones.

The data analysis strategy to observe MBH mergers will be affected by
two key factors: the event rate of mergers emitting gravitational
radiation at frequencies in the \lisa~best sensitivity window, and the
accuracy of our knowledge of the gravitational waveforms. In
Sec.~\ref{rates} we review present estimates of the event rates for
both SMBH binaries and binaries comprising one IMBH. In Sec.~\ref{wf}
we provide a list of important open problems in our knowledge of the
theoretical waveforms. For the inspiral phase these include spin
precession, eccentricity, the inclusion of high-order Post-Newtonian
terms in the amplitude and phase, and an accurate prediction of the
transition from inspiral to plunge. For the ringdown phase we point
out that numerical relativity will ultimately be required to determine
the relative quasinormal mode excitation, and to reduce the
dimensionality of the template space in matched filtering.

\section{Massive black hole binary event rates}\label{rates}

The present observational evidence for the existence of astrophysical
BHs is strong and growing \cite{narayan}.  The most convincing case
comes from observations of stellar proper motion in the center of our
own galaxy, indicating the presence of a ``dark object'' of mass
$M\simeq (3.7\pm 0.2)\times 10^6~M_\odot$ \cite{SgrA} and size smaller
than about one astronomical unit \cite{shen}. A Schwarzschild BH of
the given mass has radius $R=2GM/c^2\simeq 0.073$ astronomical units,
compatible with the observations. Any distribution of individual
objects within such a small region would be gravitationally unstable
\cite{maoz}, and theoretical candidates alternative to BHs (such as
boson stars and gravastars) are probably unlikely to exist in
nature. Furthermore there is strong observational evidence for the
presence of MBHs in the bulges of nearly all local, massive galaxies
\cite{MBHs}. These BHs have masses in the range $M\sim
10^5-10^9~M_\odot$, approximately proportional to the mass of the host
galaxies, $M\sim 10^{-3}~M_{\rm galaxy}$ \cite{merrittferrarese}.
Recent observations led to other remarkable discoveries.  There is an
almost-linear relation between the mass of a MBH and the mass of the
galactic bulge hosting the BH \cite{MBHs}. The BH mass is also tightly
correlated with other properties of the galactic bulge, such as the
central stellar velocity dispersion $\sigma$, the bulge light
concentration and the near-infrared bulge luminosity
\cite{Msigma}. Applying over an enormous mass range, these
correlations clearly indicate that MBHs are somehow aware of the
surrounding galactic environment.

Details of the formation process of MBHs are not well known. A popular
formation scenario involves the collapse of primordial, massive
($M\sim 30-300~M_\odot$) metal free Population III stars \cite{ohkubo}
at cosmological redshift $z\sim 20$ to form primordial BHs with $M\sim
10^2~M_\odot$, clustering in the cores of massive dark-matter halos
\cite{madaurees}. In some alternative scenarios, BH seeds of larger
mass $M\sim 10^5~M_\odot$ form at $z\gappreq 12$ from low-angular
momentum material in protogalactic discs \cite{KBD} (see also
\cite{BVR}). A major source of uncertainty in predicting the evolution
of MBHs comes from the unknown ``occupation number'' (fraction of
galaxies containing a MBH) at high redshifts. As dark matter halos
merge (maybe starting early, at $z\sim 20$), seed MBHs can grow both
through gas accretion (which is perhaps the dominant mechanism
\cite{volonteri1}) and through coalescence with other MBHs.

Larger galaxies grow through the agglomeration of smaller galaxies and
protogalactic fragments. If more than one of these fragments contains
a MBH, MBHs will form a bound system in the merger product
\cite{BBR}. The electromagnetic observation of a MBH binary is a hard
task, requiring the study of some emission component close to the
BHs. So far there is no completely convincing observational case for
the detection of ``hard'' MBH binaries, that is, binaries having
orbital velocities larger than the velocity dispersion of stars in the
galactic nucleus \cite{MM}.


The formation of MBHs during galaxy mergers is a challenging problem
in theoretical astrophysics. The general scenario was outlined in the
pioneering analysis of \cite{BBR}, and an excellent review of the
state of the art in this field can be found in \cite{MM}. The
evolution of a MBH binary can be roughly divided in three phases: i)
as the galaxies merge, MBHs sink to the center via dynamical friction;
ii) the binary's binding energy increases because of gravitational
slingshot interactions: the ejection of stars on orbits intersecting
the binary (these stars' angular momentum must be in a region of phase
space called the ``loss cone''); iii) if the binary separation becomes
small enough, gravitational radiation carries away the remaining
angular momentum. Notice that the gravitational wave coalescence time
is shorter for more eccentric binaries \cite{PM}, and as a result
high-eccentricity binaries could be more likely to coalesce within a
Hubble time. In Sec.~\ref{wf} we will briefly sketch the complications
in data analysis (and the advantages in terms of parameter estimation)
for binaries with non-zero eccentricity.

The transition from phase ii) to phase iii) is a field of active
research, that has been referred to as the ``final parsec problem''
\cite{MM}. Since the binary will quickly eject all stars through
gravitational slingshot interaction, the problem is to find some
mechanism to refill the loss cone. Unfortunately $N$-body simulations
of spherical galaxies do not provide very reliable answers. The reason
is that the binary's hardening rates depend strongly on $N$, roughly
decreasing as $N^{-1}$ at the largest values of $N$ allowed by present
simulations. Proposed mechanisms to overcome the final parsec problem
include gas accretion, star-star encounters and triaxial distortions
of galactic nuclei \cite{hbs}. Recent simulations show that if the
galaxy is allowed to rotate, hardening rates become independent of $N$
and binaries {\it do} coalesce within a Hubble time \cite{berczik}.

These recent results are consistent with some observational evidence
indicating that {\it efficient coalescence is the norm}. First of all,
as we mentioned earlier, at present there is no convincing evidence
for bound SMBH binaries. In galaxies containing an uncoalesced binary,
mergers would bring a third black hole into the nucleus, and the
resulting gravitational slingshot interaction would eject one or more
MBHs from the nucleus. This would produce off-center MBHs, but so far
off-center MBHs have escaped detection. Since we don't detect
off-center black holes\footnote{In \cite{hoffmanloeb} three-body
slingshot interactions were proposed to explain the observations of
the bright quasar HE0450-2958, which seems not to be surrounded by a
massive host galaxy. See however \cite{merritt-noeject} for an
alternative explanation that does not require the quasar to be
ejected.}, coalescence must proceed on short timescales. In addition,
Haenhelt \cite{HBHsMerge} remarked that MBH ejections by three-body
slingshot interactions would weaken the tight $M-\sigma$ correlations
that are observed. More speculative evidence for efficient coalescence
comes from observations of radio galaxies. About a dozen radio
galaxies exhibit abrupt changes in the orientation of their radio
lobes, producing an $X$-shaped morphology. According to some
theoretical models, the MBH producing the jet could have undergone a
spin flip, possibly produced by capture of a second MBH: perhaps in
$X$-shaped radio sources we are {\it already} observing merger events
\cite{merritt}.

The third (gravitational-radiation dominated) phase in the evolution
of MBH binaries has recently attracted a lot of attention from the
astrophysics community. The reason is that, according to general
relativity, unequal-mass binaries should radiate not only energy and
angular momentum, but also linear momentum \cite{bekenstein}. The
resulting gravitational wave recoil speed could be large enough to
kick MBHs out of the host galaxy \cite{favata}. Unfortunately, a large
fraction of this linear momentum would be radiated in the final phases
of the MBH binary coalescence, where black hole perturbation theory
and Post-Newtonian (PN) expansions of the Einstein field equations are
less reliable.  Recent PN calculations set an upper limit of $v_{\rm
kick}\sim 250\pm 50$~km~s$^{-1}$ on the resulting recoil speed
\cite{blanchetwill}. Observationally, MBH ejections by gravitational
wave recoil would produce some scattering in the $M-\sigma$
relation. Recent arguments suggest that the observed scattering
already constrains the magnitude of the kick to values $v_{\rm
kick}\lappreq 500$~km~s$^{-1}$, compatible with the general
relativistic upper limit \cite{libeskind}. In the near future further
observations may rule out gravitational wave recoil as a viable
mechanism for MBH ejection from galactic cores. Calculations assuming
the largest possible value allowed for the magnitude of the kick show
that, if viable, gravitational wave recoil would lower dynamical
friction, hence lower the rates of MBH binaries observable by \lisa~by
factors $\sim 10$ \cite{micic}.

The \lisa~noise curve determines the optimal mass and redshift range
where binary inspiral and ringdown events have large SNR, allowing a
precise measurement of the source parameters (see Fig.~\ref{BBWSNRz}
below). Reliable estimates of the number of events detectable during
the mission's lifetime, and of their mass spectrum as a function of
redshift, will play a key role in the planning of \lisa~data
analysis. For this reason, over the last few years the calculation of
MBH merger event rates and of their mass spectrum has become an active
field of research.

In a pioneering paper \cite{H94} Haehnelt noticed that the event rate
inferred from the quasar luminosity function is too low to be
detectable, but event rates can be very high if we assume that MBHs
reside in dark matter halos. Various authors have recently re-computed
these event rates. A major factor influencing their predictions is the
model used to deal with the merger history of dark halos. Menou {\it
et al.} use merger tree algorithms to show that the ubiquity of MBHs
in luminous galaxies today \cite{MBHs} is consistent with a small
``occupation number'' (fraction of galaxies containing a MBH) at high
redshifts. They predict an integrated rate of $\sim 10$~events/year
for MBH mergers out to $z\sim 5$ \cite{menou}.  Wyithe and Loeb use a
semianalytic model of dark matter halo mergers which assumes that all
halos contain MBHs, thus overestimating the event rate by about one
order of magnitude \cite{WL}. Revised estimates by Rhook and Wyithe
using a more conservative occupation number yield rates of
$15$~events/year, consistent with \cite{menou}, and suggest that most
events should originate at $z\sim 3-4$ \cite{RW}. The estimates in
\cite{menou,RW} should be considered somewhat optimistic, in the sense
that both works assume coalescence to be rapid. Sesana {\it et al.}
use a more conservative approach to coalescence: in their model for
binary evolution some binaries can be ejected from the galactic
core. Using a seeded MBH growth model, they estimate that 3 years of
\lisa~observations could resolve $\sim 35$ MBH mergers in the redshift
range $2\lappreq z\lappreq 6$. A fraction of these mergers ($\sim
9$~events/year) would contain at least one black hole heavier than
$10^5~M_\odot$ \cite{sesana1,sesana2}. Enoki {\it et al.} use a
semianalytic model of galaxy and quasar formation based on the
hierarchical clustering scenario to estimate the stochastic background
due to inspiralling MBH binaries. They find that \lisa~could detect
binaries with total mass $M<10^7~M_\odot$ and $z>2$ at a rate of
1~event/year; events with $M\sim 10^8$ would mostly be observed at
$z<1$, and events with $M\sim 10^6$ would be visible at $z\sim 3$
(though with lower amplitude) \cite{EINS}.

Other authors predict more optimistic event rates than those listed so
far. Islam {\it et al.} estimate that, if merger is efficient,
$10^4-10^5$ events/year could be observed in the \lisa~band; these
events could also be coincident with gamma-ray burst observations
\cite{ITS}.  Scenarios in which BH seeds of large mass $M\sim {\rm
a~few}\times 10^5~M_\odot$ form at $z\gappreq 12$ from low-angular
momentum material in protogalactic discs \cite{KBD} predict even
larger rates.  In fact, in these scenarios MBHs should produce a noisy
stochastic background similar to the white dwarf binary background,
but with much larger SNR \cite{KZ}.

\begin{table}[hbt]
\centering
\caption{SMBH binary rates (events/year) predicted by different models.}
\vskip 12pt
\begin{tabular}{@{}lcc@{}}
\hline
\hline
Reference                                       & Rate             & Redshift range\\
\hline
\hline
Haenhelt 2003 \cite{HBHsMerge}                  & 0.1-1            & $0<z<5$ (gas collapse only) \\
                                                & 10-100           & $z>5$ (hierarchical buildup) \\
\hline
Menou {\it et al.} 2001 \cite{menou}            & 10               & $z<5$ \\
\hline
Rhook and Wyithe 2005 \cite{RW,WL}              & 15               & $z\sim 3-4$\\
\hline
Sesana {\it et al.} 2004 \cite{sesana1,sesana2} & 35               & $2<z<6$\\
                                                & 9                & one BH with $M>10^5 M_\odot$\\
\hline
Enoki {\it et al.} 2004 \cite{EINS}             & 1                & $z>2$\\
\hline
Islam {\it et al.} 2003 \cite{ITS}              & $10^4-10^5$      & $z\sim 4-6$\\
\hline
Koushiappas and Zentner 2005 \cite{KZ}          & stochastic background & mostly $z\sim 10$, down to $z\sim 1$\\
                                                &                  & (see their Fig.~3)\\
\hline	     	   	       	            	 	     
\hline								     
\end{tabular}
\label{tab:rates}
\end{table}

Given the significant differences between MBH binary formation models
and the predicted event rates, we find it useful to provide a
schematic summary of the available literature on event rates in
Table~\ref{tab:rates}. The numbers we list should be interpreted with
caution. Each prediction depends on a large number of poorly known
physical processes, and the notion of ``detectability'' of a merger
event is defined in different ways: some authors define detectability
setting a threshold on the SNR, others set a threshold on the
gravitational wave effective amplitude. Furthermore, different authors
use different \lisa~noise curves. Some of them assume that the
low-frequency $f^{-2}$ dependence of the \lisa~acceleration noise can
be extrapolated below $10^{-4}$~Hz, which increases the event rates by
including highly redshifted, high mass BH mergers. In reality the
noise curve will probably rise very steeply below $\sim 3\times
10^{-5}$~Hz. This affects also the SNR and parameter estimation (see
\cite{BBW} for a discussion).

A tentative bottom line is that we could face one of the following two
scenarios. According to a class of models, we should observe $\approx
10$ events/year at redshifts (say) $2\lappreq z\lappreq 6$. However,
we cannot exclude the possibility that hundreds or thousands of MBHs
will produce a large (and perhaps stochastic) background in the
\lisa~data. Clearly, the detection strategy to use strongly depends on
which of the two scenarios actually occurs in nature. At this stage,
our best bet is to devise techniques of detection and parameter
estimation that encompass both possibilities.

\subsection{Binaries involving intermediate-mass black holes}

Until recently astrophysical BHs were thought to belong to either one
of two broad mass ranges: stellar-mass BHs with $M\sim 3-20~M_\odot$
(produced by the collapse of massive stars) and SMBHs with $M\sim
10^5-10^9~M_\odot$, which have been the main focus of our discussion
so far. In the last few years observations of ultraluminous X-ray
sources, combined with the fact that several globular clusters show
evidence for an excess of dark matter in their cores, provided strong
hints of the existence of astrophysical IMBHs with $M\sim
10^2-10^4~M_\odot$ (see \cite{millercolbert} for a review).

It is possible that MBH binaries have a mass ratio $q\equiv m_2/m_1$
significantly smaller than one. In fact, recent studies by Volonteri
{\it et al.} \cite{volonteri1,volonteri} suggest that low-redshift
($z\lappreq 10$) MBH mergers predominantly occur with a mass ratio
$q\simeq 0.1$ or smaller. Some of these binaries could be SMBH-IMBH
binaries. If MBH binary coalescence timescales are long enough,
three-body slingshot interactions and gravitational wave recoil may
generate a population of IMBHs wandering in galaxy halos at the
present epoch \cite{volonteriperna}.

Event rates for IMBH-IMBH binaries (that is, binaries containing a
$10-100~M_\odot$ BH orbiting a $100-1000~M_\odot$ BH) were first
estimated by Miller \cite{miller} and then revised by Will
\cite{will}. The revised estimates are very pessimistic, predicting
$\sim 10^{-6}$ events/year for typical values of the parameters.

A more promising scenario for gravitational wave detection emerges if
IMBHs play a role in the formation of SMBHs. The process can roughly
be split in three stages \cite{ebisuzaki}. In the first stage IMBHs
are formed by runaway mergers of massive stars in dense young stellar
clusters \cite{ebisuzaki}, or alternatively by runaway mergers of
smaller black holes in globular clusters \cite{millerhamilton}, which
are typically much older (so that all stars of mass $\gappreq
0.8~M_\odot$ have evolved off the main sequence, and the most massive
objects are compact remnants) \footnote{The viability of runaway
mergers in young stellar clusters is supported by numerical
simulations, while in the globular cluster scenario growth times are
rather long, making it harder to explain why young clusters (such as
MGG-11) would contain IMBHs.}.  In the second stage these IMBHs
accumulate at the galactic center due to sinkage of the clusters by
dynamical friction. In the third and final stage, the IMBHs merge
(either by successive multi-body interactions or spiralling into a
preexisting central SMBH) emitting gravitational radiation.

Matsubayashi {\it et al.} \cite{MSE} estimated the inspiral and
ringdown radiation emitted in the formation of a $10^6~M_\odot$ SMBH
by merging of a thousand IMBHs of mass $10^3~M_\odot$, following two
radically different merging histories. In the {\it hierarchical}
scenario pairs of equal-mass BHs merge producing a single, more
massive BH, and then the process repeats. Conversely, in the {\it
monopolistic} (or runaway) scenario, a single BH grows through
subsequent mergers with surrounding BHs. In the hierarchical scenario
the majority of mergers occurs between low-mass black holes, so most
of the radiation is emitted in the \lisa~high frequency band (or even
at frequencies $f\gappreq 10~$Hz, too large to be detected by
\lisa). In the monopolistic scenario, many events occur when the mass
of the accreting black hole is large enough for the radiation to be in
the optimal \lisa~band. Event rates are extremely uncertain and depend
in a similar way on the merging history. In the monopolistic scenario
they could be as high as $~20-70$ events/year if all galaxies
experience mergers of $\sim 10^3~M_\odot$ BHs. In the hierarchical
scenario, given the poor sensitivity of \lisa~at high frequencies,
they could be smaller by about one order of magnitude\footnote{Seto
{\it et al.}  \cite{decigo} proposed to shorten the \lisa~armlength by
a factor $\sim 10$. The best sensitivity region of the resulting
instrument ({\it DECIGO}) would be shifted towards higher frequencies,
dramatically improving the event rates in the hierarchical
scenario. The price to pay is that {\it DECIGO} would have low
sensitivity to gravitational waves from BH binaries of mass $\gappreq
10^5~M_\odot$.}.

A detailed $N$-body simulation of the sinking of a $3000~M_\odot$ IMBH
into a $6\times 10^6~M\odot$ SMBH can be found in \cite{MME}.
Dynamical friction becomes ineffective at the orbital radius inside
which the initial stellar mass is comparable with the IMBH mass. Quite
surprisingly, simulations show that at this point the IMBH's orbital
eccentricity grows, lowering the merging timescale due to
gravitational radiation. Variants of this scenario have been
considered by Miller and by Portegies-Zwart {\it et al.}
\cite{miller2,portegies}. Miller estimated a detection rate of a few
events/year, and suggested that mergers of a $10^3~M_\odot$ IMBH into
a $10^6~M_\odot$ SMBH could be observed out to $z\sim 20$ at SNR 10 in
a one-year integration. Typical SNRs could be much larger than the
typical SNRs for EMRIs ($\sim 10^3$ instead of $\sim 10$), allowing
precise parameter estimation \cite{miller2} (but see below for
theoretical issues in modelling the waveforms of SMBH-IMBH
binaries). Portegies-Zwart {\it et al.}  predict an even more
optimistic rate of $\sim 10^2$ events/year throughout the universe
\cite{portegies}. These estimates are very preliminary and even more
uncertain than the corresponding estimates for SMBH binaries, but they
should be taken into account to decide (for example) the optimal
armlength of~\lisa.

\section{Open problems in waveform modelling}\label{wf}

In the last decades inspiral waveforms for circular orbits, which are
crucial for the detection of gravitational waves by ground-based
interferometers, have been studied in depth. Expansions of the phasing
are known up to 3.5PN order if spin terms are negligible, and up to
2PN order for binaries with spins aligned (or antialigned) and normal
to the orbital plane. Leading-order contributions to the phasing from
alternative theories of gravity (scalar-tensor theories and theories
allowing for a non-zero graviton mass) have also been studied in the
context of \lisa~(see \cite{BBW,BBW2} for an extensive discussion).

The measurability of ringdown waves with \lisa~has been studied in
\cite{BCW}. For a Schwarzschild BH the real part of the fundamental
quasinormal mode frequency $\omega=2\pi f+\ii/\tau$ is in the optimal
region of the \lisa~sensitivity curve, $f=1.207\times 10^{-2}(10^6
M_\odot/M)$, and the damping time $\tau=55.37(M/10^6 M_\odot)$ is
slightly larger than the light travel time across one of the
\lisa~arms $T_{\rm light}=(5\times 10^9~{\rm m})/c\simeq 17$~s.

\begin{figure*}
\begin{center}
\epsfig{file=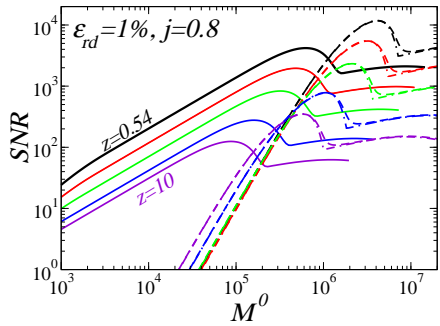,width=5cm,angle=-90}
\caption{SNR for the last year of inspiral as a function of the total
mass of the binary (continuous lines) and SNR for ringdown as a
function of the BH mass (dashed lines). In both cases the masses $M^0$
are evaluated in the source frame: the mass measured at the detector
is $M=(1+z)M^0$. For the ringdown we pick the fundamental $l=m=2$
quasinormal mode frequency with specific angular momentum $j=0.8$,
assuming an efficiency $\epsilon_{\rm rd} = 1 \%$.  From top to bottom
the lines correspond to redshifts $z=0.54$, $z=1$, $z=2$, $z=5$ and
$z=10$ (from \cite{BCW}).
\label{BBWSNRz}}
\end{center}
\end{figure*}

The main uncertainty affecting the ringdown SNR concerns the ringdown
efficiency $\epsilon_{\rm rd}$ (ratio of energy radiated to the BH
mass), plausible values of which range perhaps between $0.1\%$ and
$3\%$. The effect of angular momentum is less pronounced
\cite{BCW}. In Fig.~\ref{BBWSNRz} we plot the SNR for observations of
the last year of inspiral of equal-mass BH binaries (as a function of
the binary's total mass in the source frame), and compare it with the
SNR for the ringdown of the finally formed BH (as a function of its
mass in the source frame). We compute this quantity for different
values of the cosmological redshift, assuming a cosmology with
$\Omega_M=0.3$, $\Omega_\Lambda=0.7$ and $H_0=0.72$. The inspiral and
ringdown SNRs are comparable. For example, at $D_L=3~$Gpc ($z\simeq
0.5$) the ringdown SNR can be as large as $\sim 10^4$ for BHs of mass
$\sim 4\times 10^6~M_\odot$, and the inspiral SNR can be as large as
$\sim 4\times 10^3$ for a total mass of the binary $\sim 7\times
10^5~M_\odot$.

Since the SNR is so large and errors on the source parameters scale
with the inverse of the SNR, \lisa~can provide very accurate
measurements of the source parameters both for inspiral
\cite{BBW,BBW2} and for ringdown \cite{BCW}. This suggests the
exciting possibility to measure (say) the masses of the BHs in a
binary, or the mass of the BH they form after merger, by matched
filtering of the inspiral and ringdown gravitational waveforms,
respectively. A problem with this idea is that \lisa~does not measure
masses in the source frame $M^0$, but only a redshifted combination
$M=(1+z)M^0$. A possibility to disentangle the $z$-dependence is to
measure the luminosity distance $D_L(z,\Omega_M,\Omega_\Lambda,H_0)$
and (assuming that cosmological parameters are known to a good
accuracy) invert this relation to get
$z(D_L,\Omega_M,\Omega_\Lambda,H_0)$ \cite{hughes}. In the range where
we expect most events (say $2\lappreq z\lappreq 6$) the error on $D_L$
is rather small, typically less than $\sim 10\%$: see eg. Fig.~7 in
\cite{BBW}. By the time \lisa~flies, weak lensing errors can be
expected to dominate over other sources of error
\cite{holzhughes,kocsis}. If SMBH mergers are accompanied by gas
accretion leading to Eddington-limited quasar activity, and if spin
precession reduces the errors in parameter estimation, the \lisa~error
volume may be small enough to contain a single quasar out to $z\sim 3$,
allowing a test of the hypothesis that gravitational wave events are
accompanied by bright quasar activity \cite{kocsis}.

This ambitious program relies on a detailed knowledge of the
gravitational waveforms, which is necessary to reduce errors in
parameter estimation. Our present knowledge of theoretical templates
should be extended to include the following effects:

{\it 1) Spin precession - } In \cite{BBW} we observed that including
spin-orbit and spin-spin terms in the gravitational wave phasing
degrades the accuracy of parameter estimation. This is easy to
understand, since the spin-orbit and spin-spin parameters are strongly
correlated with other parameters in the phasing (such as the masses)
and adding more parameters effectively dilutes the available
information. However in \cite{BBW} we considered only spins aligned
(or antialigned) and normal to the orbital plane, which is probably
not realistic. In general the relativistic spin-orbit interaction
causes the orbital plane to precess in space, producing a
characteristic modulation of the waveforms. A preliminary analysis by
Vecchio shows that, for a $(10^6+10^6)~M_\odot$ binary, the additional
information coming from spin precession can improve parameter
estimation by factors $\sim 3-10$ in angular resolution and luminosity
distance, and by factors $\sim 10$ and $\sim 10^3$ in chirp mass and
reduced mass, respectively \cite{vecchio}. A more general and
systematic analysis is urgently needed.

{\it 2) Eccentricity - } For Earth-based interferometers, neglecting
the orbital eccentricity of a binary is an excellent
approximation. Earth-based detectors can only observe binaries in the
very last stages of the inspiral, when radiation reaction has had
enough time to circularize the orbits \cite{PM}. For \lisa~the
situation is different. As we discuss below, MBH binaries could have a
significant eccentricity in the last year of inspiral, especially if
their mass ratio is not close to one (which is the case for SMBH-IMBH
binaries). It is well known that orbital eccentricity produces
radiation at all harmonics of the Keplerian frequency $f_K$: $f_{GW}=n
f_k$ ($n=1,~2,~3\dots$) \cite{PM}.  For high-mass SMBH binaries the
gravitational wave frequency for circular orbits $f_{GW}=2 f_K$ could
be too low to be detected by \lisa~with high SNR, but higher harmonics
may be in the optimal frequency band. In addition, the presence of
higher harmonics in the signal effectively provides more information
on the source, improving parameter estimation \cite{hellings}.

Analytical calculations and $N$-body simulations show that, in purely
collisionless spherical backgrounds, the expected equilibrium
distribution of eccentricities is skewed towards high $e\simeq
0.6-0.7$, and that dynamical friction does not play a major role in
modifying such a distribution (\cite{colpi}, in particular
Fig.~5). Recent simulations using smoothed particle hydrodynamics
follow the dynamics of binary BHs orbiting in massive, rotationally
supported circumnuclear discs \cite{dotti}. The rotation of the disc
circularizes the orbit if the binary {\it corotates} with the disc,
possibly increasing the merging timescale due to gravitational
radiation so much that the binary stalls and no coalescence
results. If the orbit is {\it counterrotating} the initial
eccentricity does not decrease, and BHs may enter the gravitational
wave emission phase with high eccentricity. For corotating discs, the
numerical resolution of these simulations is not good enough to
compute the precise value of the (small) residual eccentricity when
the BHs are close enough that gravitational radiation takes over.

Complementary studies show that eccentricity evolution may still occur
in later stages of the binary's life, because of close encounters with
single stars \cite{BMS} and/or gas-dynamical processes
\cite{armitage}. Stellar dynamical hardening might leave the binary
with non-zero eccentricity, although most studies suggest that any
such eccentricity would be small \cite{BMS} (see however \cite{MME}
and \cite{aarseth} for recent examples of eccentricity growth in
$N$-body simulations). On the other hand, the gravitational
interaction of the binary with a surrounding gas disc is likely to
increase the eccentricity of BH binaries. The transition between
disc-driven and gravitational wave-driven inspiral can occur at small
enough radii that a small but significant eccentricity survives,
typical values being $e\sim 0.02$ (with a lower limit $e\simeq 0.01$)
one year prior to merger (cf. Fig.~5 of \cite{armitage}). If the
binary has an ``extreme'' mass ratio $q\lappreq 0.02$ the residual
eccentricity can be considerably larger, $e\gappreq 0.1$.

\begin{figure*}
\begin{center}
\epsfig{file=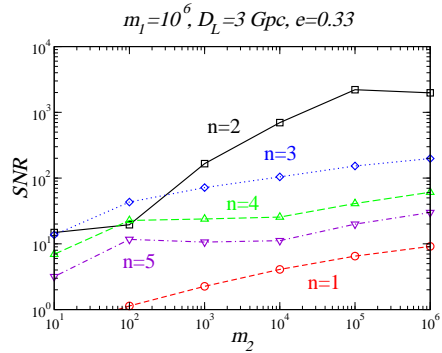,width=5cm,angle=-90}
\caption{{\it (Courtesy of Jim Shifflett)} SNR for different harmonics
of an eccentric binary at luminosity distance $D_L=3$~Gpc, observed
during the last year of inspiral. We assume that the eccentricity
$e=0.33$ one year prior to merger. The more massive BH has
$m_1=10^6~M_\odot$, and we plot the SNR of different harmonics as a
function of the lighter BH's mass $m_2$ (in $M_\odot$).
\label{e33}}
\end{center}
\end{figure*}

For concreteness, in Fig.~\ref{e33} we show the SNR for different
harmonics of an eccentric binary at luminosity distance $D_L=3$~Gpc,
observed during the last year of inspiral. We assume that the
eccentricity $e=0.33$ one year prior to merger. The more massive BH
has $m_1=10^6~M_\odot$, and we plot the SNR of different harmonics as
a function of the lighter BH's mass $m_2$ (in $M_\odot$). Harmonics
with $n=2$ to $n=5$ are detectable for all values of the mass
ratio. For low mass ratios $q\lappreq 0.01$ (that is, for SMBH-IMBH
binaries and EMRIs) the SNR of the $n=3$ harmonic can be comparable
with the SNR of the $n=2$ harmonic. A more extensive survey of the
parameter space is clearly needed \cite{BSW}. For SMBH-IMBH binaries,
including eccentricity could be necessary for detection. For SMBH
binaries it will be very useful for detection of mergers in the
high-mass end ($M\gappreq 10^7~M_\odot$), and possibly important for
parameter estimation.

{\it 3) High-order PN effects in phasing and amplitude - } The
\lisa~SNR for MBH inspirals can be very large (see
Fig.~\ref{BBWSNRz}). When the SNR is so large we must take into
account the possibility that {\it systematic} errors, due to the
truncation of the phasing at some given PN order, could be comparable
with {\it statistical} errors. Preliminary results \cite{BBC} show
that the contribution of high-order PN terms in the phasing is not
particularly significant for equal-mass mergers, but can be very
relevant when the mass ratio is small. As a simple measure of the
convergence properties of the PN expansion we can compute the number
of cycles from different PN contributions. For a $(10^6+10^6)~M_\odot$
binary, the Newtonian, 1PN, 2PN and 3PN terms contribute $(\sim
2300,\sim 100,\sim 5,\sim 2)$ cycles, respectively. As the mass ratio
decreases the PN expansion gets worse. For a $(10^6+10^5)~M_\odot$
binary the relative contributions are $(\sim 5000,\sim 300,\sim 10,
\sim 2)$ cycles, and for a SMBH-IMBH binary of $(10^6+10^3)~M_\odot$
they become $(\sim 27000,\sim4000,\sim 400,\sim 30)$ \cite{BBC}.

The bottom line is that for SMBH-IMBH binaries, not only high-order
harmonics have large relative SNR if $e\neq 0$: high-order PN terms
contribute many cycles even for zero eccentricity.  Since the PN
approximation becomes inaccurate for these systems, one could think
about using the Teukolsky formalism (based, roughly speaking, on an
expansion in $q$ keeping only the linear term). However for SMBH-IMBH
binaries the mass ratio can be rather large ($q\gappreq 10^{-3}$), and
the accuracy of the Teukolsky formalism becomes questionable. This
``buffer zone'' between EMRIs and SMBH binaries calls for the
development of a hybrid approach, taking the best from both the PN
expansion and the Teukolsky formalism.

Preliminary results show that the usual ``restricted PN
approximation'' (where the amplitude is computed using the quadrupole
formula, and only PN corrections in the phasing are considered) may be
inappropriate for \lisa. The inclusion of leading-order PN corrections
to the {\it amplitude} may be necessary for detection, even for SMBH
binaries with moderate mass ratios $q\sim 0.1$ \cite{porter}.

{\it 4) Transition from inspiral to plunge - } For ground-based
interferometric detectors, SNR estimates suggest that the first
detection may concern BH binaries of total mass $M\gappreq
25~M_\odot$. In this case the most useful part of the waveform is
emitted in the last $\sim 5$ orbits of the inspiral and during the
plunge, that takes place after crossing the last stable orbit
\cite{DIS}.  The transition from inspiral to plunge is not so
important for MBH observations with \lisa, since typically we should
be able to observe thousands of cycles of inspiral. However, if the
\lisa~acceleration noise is not under control below (say) $\sim
10^{-4}$~Hz, knowledge of the transition from inspiral to plunge could
be important to detect high-mass ($M\gappreq 10^7~M_\odot$)
binaries. An accurate model of this phase would also be useful to
predict the initial conditions of the merger, eventually leading to an
estimate of the relative quasinormal mode excitation in the ringdown
\cite{BCW2}. Most importantly, the transition from inspiral to plunge
is crucial to estimate the kick velocity. The reason is that (as
pointed out in \cite{blanchetwill}) the maximum velocity accumulated
in the inspiral phase is $\sim 20$~km~s$^{-1}$, so that the largest
contribution to the kick comes from the plunge.

{\it 5) Merger and ringdown waveforms - } Ringdown waveforms are
linear superpositions of damped exponentials whose frequencies $f$ and
damping times $\tau$ (or equivalently, quality factors $Q=\pi f \tau$)
are well known. The main uncertainty here comes from our poor
knowledge of the merger phase in generic situations (black holes with
different masses, spins, spin orientations etcetera), which in turn
affects our knowledge of the energy distribution between different
modes (see \cite{BCW} and \cite{BCW2} for a detailed
discussion). Numerical relativity will ultimately tell us which
angular components (more technically: which values of $(l,m)$ in the
spin-weighted spheroidal harmonic decomposition of the perturbations)
are excited in a realistic merger. This is an important issue, since
it will determine whether we can use \lisa~to test the no-hair theorem
through observations of the ringdown \cite{BCW}.

Creighton \cite{creighton} provides a simple estimate of the number of
filters needed for detection of (single-mode) ringdown signals for
ground-based detectors. His estimate carries over directly to the case
of \lisa. Assuming that the noise power spectrum is approximately
constant over the frequency band of two neighbouring filters, and that
our template bank in the ($f,~Q$) plane covers the range $0\leq Q\leq
Q_{\rm max}$, $f_{\rm min}\leq f\leq f_{\rm max}$, the minimum number
of filters we need is
\bea
N_{\rm filters}&\sim& 
\f{1}{4\sqrt{2}}(ds^2_{\rm max})^{-1}Q_{\rm max}
\ln\left(\f{f_{\rm max}}{f_{\rm min}}\right)
\simeq 1085
\left(\f{Q_{\rm max}}{20}\right)
\left(\f{ds^2_{\rm max}}{0.03}\right)^{-1} \nonumber \\
&\times&\left\{1+
\f{1}{\log 10^4}
\left[\log\left(\f{f_{\rm max}}{1~{\rm Hz}}\right)
-\log\left(\f{f_{\rm min}}{10^{-4}~{\rm Hz}}\right)\right]
\right\}
\,.\nonumber
\eea
%
Here $ds^2$ is a metric measuring distances in the template space
\cite{owen}, and a maximum distance $ds^2_{\rm max}=3 \%$ corresponds
to losing $10 \%$ of the expected event rate due to a mismatched
template.  For detection of $N$ single-mode waveforms, an estimate of
the filters we need can be obtained multiplying the previous number by
$N$. The problem of determining the number of filters required to {\it
resolve} two or more modes \cite{BCW} has not been discussed so
far. Clearly, knowing {\it which} modes are excited in a realistic
merger can dramatically reduce the dimensionality of the template
space.



\section{Conclusions and outlook}\label{conclusions}

The observation of MBH mergers is potentially the most rewarding
\lisa~milestone. The data analysis strategy will depend on two key
elements: the event rate of mergers emitting gravitational radiation
at frequencies in the \lisa~best sensitivity window, and the accuracy
of our knowledge of the gravitational waveforms.

In Table \ref{tab:rates} we provide a schematic summary of event rate
estimates. As testified by the spread between different models, these
estimates depend on a large number of poorly known physical
processes. In making comparisons we must also account for the fact
that different authors use different notions of ``detectability'' of a
merger event and different models of the \lisa~noise curve (event
rates are particularly sensitive to the assumed \lisa~acceleration
noise below $10^{-4}$~Hz).
Roughly speaking, we could face one of the following two scenarios.
According to a class of models, we should observe $\approx 10$
events/year at redshifts $2\lappreq z\lappreq 6$. In the second class
of models, hundreds or thousands of MBHs will produce a large (and
perhaps stochastic) background in the \lisa~data: to borrow the
terminology used for the white dwarf binary background, this would be
a very noisy ``cocktail party problem''. We should be ready for any of
the two scenarios to actually occur in nature, devising techniques of
detection and parameter estimation that encompass both possibilities.

The physics we can learn from the observations will depend on our
ability to develop reliable theoretical templates for the
waveforms. For the inspiral phase, the most important effect should be
the strong modulation in the waveforms produced by spin precession. An
important difference with Earth-based detectors is that (depending on
the binary's formation process) the inspiral signal could have
significant residual eccentricity when it enters the
\lisa~band. Furthermore, given \lisa's potentially large SNR,
higher-order PN effects in phasing and amplitude must be included to
improve parameter estimation. Eccentricity and higher-order PN
corrections are more relevant for small mass ratios (see
eg. Fig.~\ref{e33}), and their inclusion will be necessary in the data
analysis of SMBH-IMBH binaries. An accurate model of the transition
from inspiral to plunge should only be necessary for high-mass
binaries if the MBH mass spectrum as a function of redshift is such
that a significant number of events have frequencies below
$10^{-4}$~Hz, {\it and} if \lisa~design choices do not guarantee
complete control of the acceleration noise in this frequency band.

SMBH-IMBH binaries are promising sources, despite their largely
uncertain event rates. They present us with a different challenge,
stretching the applicability limits of both PN expansions and black
hole perturbation theory. A pragmatic data analysis approach for these
systems could adopt hybrid approximation schemes, similar to the
``kludge'' waveforms for EMRIs (see S. Drasco's contribution to these
proceedings).

Simulations of the merger in numerical relativity will be useful to
assess the astrophysical significance of gravitational wave recoil.
They will also determine which modes are dominant in the ringdown
phase, reducing the dimensionality of the template space required for
ringdown detection and for tests of the no-hair theorem. Reliable
estimates of the energy and angular momentum emitted in ``generic''
merger conditions (where binary members have arbitrary mass, spin
magnitude and orientation) will provide valuable information to match
inspiral and ringdown waveforms, significantly improving parameter
estimation from both phases.

The detection of gravitational waves from MBH mergers is a challenging
interdisciplinary task, and it could open an exciting new era for
astronomy. We should be ready for surprises.

\subsection*{Acknowledgments}
I am grateful to Matt Benacquista, Alessandra Buonanno, Vitor Cardoso,
Kristen Menou, Edward Porter, Massimo Tinto, Michele Vallisneri, Cliff
Will and an anonymous referee for suggestions. I thank Jim Shifflett
for kindly allowing me to include his preliminary results on eccentric
binaries. This work was supported in part by the National Science
Foundation under grant PHY 03-53180.

\end{document}